**Electromagnetic-dual metasurfaces for topological states along a one-dimensional interface**

*Dia'aaldin J. Bisharat\* and Daniel F. Sievenpiper*

Electrical and Computer Engineering Department
University of California San Diego
La Jolla, California, 92093, USA

E-mail: dbisharat@eng.ucsd.edu

The discovery of topological insulators has rapidly been followed by the advent of their photonic analogues, motivated by the prospect of backscattering-immune light propagation. So far, however, implementations have mainly relied on engineering bulk modes in photonic crystals and waveguide arrays in two-dimensional systems, which closely mimic their electronic counterparts. In addition, metamaterials-based implementations subject to electromagnetic duality and bianisotropy conditions suffer from intricate designs and narrow operating bandwidths. Here, it is shown that symmetry-protected topological states akin to the quantum spin-Hall effect can be realized in a straightforward manner by coupling surface modes over metasurfaces of complementary electromagnetic responses. Specifically, stacking unit cells of such metasurfaces directly results in double Dirac cones of degenerate transverse-electric and transverse-magnetic modes, which break into a wide non-trivial bandgap at small inter-layer separation. Consequently, the ultrathin structure supports robust gapless edge states, which are confined along a one-dimensional line rather than a surface interface, as demonstrated at microwave frequencies by near field imaging. The simplicity and versatility of the proposed approach proves attractive as a tabletop platform for the study of classical topological phases, as well as for applications benefiting the compactness of metasurfaces and the potential of topological insulators.

**1. Introduction**

The discovery of topological insulators[1,2] in condensed-matter systems has promoted extensive research on analogous systems of classical waves including acoustics[3–7] and photonics.[8–10] Uniquely, these systems have insulating bulk but conducting interfaces that host chiral one-way[11–15] or helical spin-polarized[16–31] edge states. As a result, wave propagation is immune to backscattering unlike in ordinary photonic circuitry, where realistic fabrication imperfections, disorder or arbitrary bends could severely reduce signal transmission, hence device performance. The key to this topological protection is interfacing two claddings of different Chern numbers,[32] which characterize the geometrical phase accumulation[33] of the wavefunctions in the reciprocal space. Such an extra phase results in a distinct evolution of the field profiles and it is invariant by any gauge transformation in both broken and conserved time-reversal symmetry-based systems.

Successful photonic implementations based on breaking time-reversal symmetry include using an external magnetic field and gyromagnetic materials in the microwave regime.[11,12] An alternative approach that relies on time modulation for generating a synthetic gauge field[13] with an arbitrary spatial distribution was also demonstrated at near-infrared frequencies using coupled waveguides.[14] However, these systems, which are analogues of quantum Hall topological insulators, are notably bulky and hard to implement in an integrated platform.[15] In the case of time-reversal-invariant systems, a non-periodic array of coupled ring resonators was originally proposed, in which the spin degree of freedom is emulated by the rotation of the wave in the rings in clockwise or counter-clockwise directions. [16–18] However, the conservation of pseudospins, which relies on the directional coupling between the rings, necessitates a large ring diameter, hence an extensive footprint of the system. Alternatively, crystalline symmetry could be exploited in photonic crystal structures to give rise to modal degeneracies that play the role of pseudospin states. [19–23] Commonly, the lattice valleys are folded onto the center of the Brillouin zone of a new lattice, thus creating photonic 'molecules' supporting hybridized dipolar and quadrupolar circularly polarized eigenmodes. Meanwhile, the expansion (shrinking) of constituent elements in the unit cell causes bandgap opening with non-trivial (trivial) bulk bands. However, the interface of such systems may break the symmetry responsible for the topological order, thus causing the associated topological edge states to become gapped.

A more attractive approach for realizing photonic analogues of the quantum spin-Hall effect (QSHE) is based on the internal symmetry of electromagnetic fields (i.e. EM duality).[24–31] Here, metamaterials-based unit cells with equal permittivity ($\varepsilon$) and permeability ($\mu$) tensors are used as building blocks for a two-dimensional superlattice. The $\varepsilon/\mu$ matching condition enables constructing degenerate spin states out of linear combinations of transverse magnetic (TM) and transverse electric (TE) waves, while a bianisotropic coupling between the TE and TM modes plays a role analogous to that of spin-orbit interaction in electronic topological systems and produces a synthetic gauge field that acts separately on the decoupled spin states. However, due to the intrinsic dispersion of bianisotropic metamaterials, demonstrations in both metallic[25–27] and dielectric[28,29] structures at microwave frequencies satisfy the operation conditions only in a narrow frequency range. Furthermore, engineering such topological systems is not straightforward[29,30] and may require cumbersome designs that may be hard to realize in practice.[31]

In this work, we demonstrate an ultrathin photonic topological insulator (PTI) based on EM duality, wherein stacking open-boundary metallic metasurfaces (also called designer/spoof surface plasmon crystals) of complementary EM properties naturally cross couples between TM and TE modes. Straightforwardly, the essential modal degeneracies are formed at the high-symmetry K/K′ points of the band structure due to the use of hexagonal unit cells, while the strong effective magneto-electric coupling inherent to the overlapped metasurfaces opens a topological non-trivial bandgap over wide frequency range. Importantly, unlike the majority of existing PTIs, including those with finite thickness,[22,26,27,29] the topological phases here arise due to engineering surface waves rather than bulk waves. Consequently, the ensued topological edge modes, which occur at the boundary of such system, are confined along a one-dimensional (1D) line[34–38] rather than a two-dimensional (2D) surface interface. In addition, we present a proof-of-concept in the microwave regime and experimentally show backscattering-immune propagation of the gapless edge modes around sharp corners by direct imaging of the near field. Owing to the simplicity, compactness, tunability, and open-boundary nature of the proposed system, it constitutes an attractive platform for photonic topological applications.

## 2. Results

### 2.1. EM-Dual Metasurface and Topological Bandgap Opening

Under EM duality symmetry, whereby electric and magnetic fields are treated on equal footing, it is possible to construct pseudo-spin states in the form of orthogonal polarizations.[39] Indeed, under duality transformation of Maxwell's equations, swapping electric and magnetic terms is equivalent to changing one wave polarization for its orthogonal counterpart, indicating that both polarizations propagate similarly.[40] However, except for a vacuum medium, this is not normally the case since unequal ε and μ tensors results in unequal propagation constants for TM and TE polarizations. Nonetheless, it may be feasible to counterbalance this unequal EM response in natural materials if we consider a combination of two periodic media, which act separately on the EM modes in a dual manner, i.e. exhibit equal but opposite effects on the EM fields. Specifically, we employ a patch-type and an aperture-type metallic metasurfaces, as shown in **Figure 1**, which have a capacitive and an inductive complementary responses, hence support low-order TE and TM propagating surface modes, respectively, with identical wavenumbers $q_{TE} = q_{TM} = |\boldsymbol{q}|$.[34] The thin complementary structures are constructed according to Babinet's principle simply by interchanging the holes and the metal.[41,42] TM/TE waves here are characterized as having non-vanishing electric/magnetic field components along the z (normal) direction, respectively. Additionally, the metasurfaces are arranged in free-space such that they satisfy a mirror reflection symmetry of the form $\varepsilon(x,y,z) = \mu(x,y,-z)$.[39, 43]

To attain an intuitive understanding of the proposed system, we first consider the two metasurface layers separately. As the band structures shown in Figure 1 indicate, the respective triangular lattices comprising each of the two metasurfaces exhibit identical mode dispersion. Moreover, due to the $C_{6v}$ symmetry of the employed hexagonal unit cells, the two sets of independent TE and TM modes exhibit Dirac cones at the K/K′ points of the first Brillouin zone. Note that the degeneracy of each pair of bands is warranted (i.e. does not require fine-tuning), and the Dirac cones necessarily appear in pairs at each valley of the momentum space. Since the dispersion curves of these modes are below the light line (represented by the grey dashed line), the EM fields are tightly confined near the metallic surface in the z-direction. Accordingly, we can stack the two metasurface layers with a finite separation distance (larger than the lattice's period constant), at which the respective EM fields are non-interacting, to constitute one system, which has a four-fold TE/TM degeneracy at the Dirac point. This degeneracy effectively restores the dual symmetry responsible for the topological properties in our system.

Next, we consider the compound metasurface in the case of small inter-layer separation, where the surface modes of each layer interact with each other. This is naturally accompanied by the cross-coupling between the TM and TE modes which introduces an effective bianisotropy to the system. Then, the Maxwell's equations have the form of $\nabla \times \boldsymbol{E} = i\omega\left(\mu_0 \mu \boldsymbol{H} + \chi \boldsymbol{E}\right)$ and $\nabla \times \mathbf{H} = -i\omega\left(\varepsilon_0 \mu \mathbf{E} + \chi \mathbf{H}\right)$, where $\mu = \varepsilon$ is assumed and $\chi$ is an effective magneto-electric coefficient tensor with non-zero elements $\chi_{xy} = -\chi_{yx}$. Considering the in-phase and out-of-phase combinations of the TE and TM modes ($\psi^{\pm} = \sqrt{\varepsilon_0}\mathbf{E} \pm \sqrt{\mu_0}\mathbf{H}$) gives

$\nabla \times \left[\left(\mu \mp \chi\right)^{-1} \nabla \times \psi^{\pm}\right] = \left(\omega/c\right)^2 \left(\mu \pm \chi\right) \psi^{\pm}$.[30] This set of eigenmodes forms our two decoupled spin states ($\psi_x^-, \psi_y^-, \psi_z^+$) (spin-up) and ($\psi_x^+, \psi_y^+, \psi_z^-$) (spin-down), which are doubly degenerate in the case of $\chi = 0$. On the other hand, finite $\chi$ value lifts the spin degeneracy at the K/K' points and opens a complete bandgap throughout the Brillouin zone, while conserving the spin values for lower and upper bands (marked with solid lines in Figure 1c).

To verify the topological character of the new states, we numerically calculate the spin Chern number of the bands. This is done by evaluating the Berry curvature using the expression $\Omega^{\pm}(\mathbf{k}) = \nabla_{\mathbf{k}} \times \langle \psi_z^{\pm} | i\nabla_{\mathbf{k}} | \psi_z^{\pm} \rangle \cdot \hat{z}$, from which the spin Chern number is computed as $C^{\pm} = (1/2\pi) \int_{BZ} \Omega^{\pm}(\mathbf{k}) d^2\mathbf{k}$. Note that it is sufficient to carry out the integration about the points of broken degeneracy (i.e. K/K') since the topological properties of the band structure arise from the hybridization of modes near these points. As indicated in Figure 1c, the lower (upper) modes are characterized by $C^+ = 1$ ($-1$) and $C^- = -1$ ($1$). This implies that the formed bandgap is topologically non-trivial, thus qualifying the metasurface as a PTI. In addition, it is found that the sign of the spin Chern number becomes opposite when the orientation of the two constituent layers of the metasurface is flipped along the z-direction. This could be understood as a result of opposite signs of magneto-electric coupling of the two inverted metasurfaces, causing band inversion mechanism to the respective band structures.

## 2.2. Spin-polarized Topological Edge States

The key property of topological structures that distinguishes them from their topologically trivial counterparts is their ability to support propagating modes along the boundary while being excluded from the bulk. In particular, edge states emerge between topologically distinct structures in which the (spin-) Chern number changes across the interface. First, we consider an interface between two metasurfaces with inverted structures (i.e. reversed bianisotropy) as shown schematically in Figure 2. Conveniently, the two adjacent topological domains share a common bandgap of the bulk. According to the bulk-boundary correspondence, the difference in the magnitude of the spin-Chern numbers across the domain wall (i.e., $\left|C_L^+ - C_R^+\right| = 2$, $\left|C_L^- - C_R^-\right| = 2$, where the subscripts $L$ and $R$ denote left and right PTIs) specifies the existence of a total of four gapless edge states, two with spin-up and two with spin-down, respectively.[25] Since the two decoupled spin states are linked by time-reversal symmetry requirement, the state $\psi^+$ ($\psi^-$) with wavevector +**q** has a counterpart $\psi^-$ ($\psi^+$) with wavevector −**q** at the same frequency. Hence, the pair $\psi^+/\psi^-$ span the bandgap near the K valley as well as near K′ valley of the Brillouin zone.

The counter propagating edge states for K and K′ valleys manifest the spin-momentum locking feature of QSHE, in other words spin-chirality in the absence of inter-spin scattering. As shown in the full-wave numerical simulations in Figure 2c, the selective excitation of $\psi^+/\psi^-$ using proper combination of electric and magnetic dipole point sources ($\sqrt{\varepsilon_0}E_z = \sqrt{\mu_0}H_z$ or $\sqrt{\varepsilon_0}E_z = -\sqrt{\mu_0}H_z$) proves the pseudo-spin configuration to be uniquely defined by the direction of the mode wavevector ($q$). In addition, as indicated by the power flux within the hexagonal unit cells in the vicinity of the interface in Figure 2b, the spatial profile of the forward propagating $\psi^+$ mode is clockwise-rotating (counter-clockwise) in the left (right) region. Meanwhile, this sense of rotations is reversed for the backward-propagating $\psi^-$ mode. This apparent linkage between the spin state and the orbital state in the metasurface emulates spin-orbit coupling in electronic QSHE systems. [25] Furthermore, the electric and magnetic field distributions over the cross section of the interface waveguide, which are plotted in Figure 2d, show that the edge mode is concentrated along the line intersection of the thin metasurface films and decays rapidly into the surrounding bulk/surfaces. Uniquely, this characterizes the edge states as 1D line waves[34,35] in contrast to the typical 2D surface edge states in previously reported PTIs.

Edge states can also emerge at the boundary between topologically nontrivial and trivial systems. Hence, we expect to observe the edge modes on the external boundary (side edges) of the dual-metasurface with air. However, in contrast to electronic systems, the free-space domain does not possess a bandgap as its spectrum is filled with the electromagnetic continuum. Consequently, such modes are not confined and could readily scatter into to the radiative continuum by arbitrary perturbations. To reduce this leakage, we consider introducing a bandgap into the external boundary by placing the metasurface approximately next to another identical metasurface, as illustrated in Figure 3a. The two dual-metasurfaces are on the same level (z-coordinate) and are separated from each other by a small slot. Moreover, the metasurfaces are made completely flat (zero thickness) by diminishing the distance between the associated inductive and capacitive metallic layers. This configuration supports two pairs of $\psi^+/\psi^-$, similar to the configuration in Figure 2a, with a pair at each edge across the slot. Here, however, the spin states are excited bi-directionally, as shown in Figure 3c, as the reversal of the air-metasurface interface orientation across the slot entails the reversal of the spin states' propagation direction at the respective edges of the two metasurfaces.

Note that the dimensions of the constituent patch and aperture of the unit cells in the case of the flat dual-metasurface are slightly altered to avoid their spatial overlap. While this makes the inductive and capacitive layers not have the exact band dispersion (i.e. not have perfect EM duality), it is found that the compound metasurface remains topologically nontrivial. As can be seen from Figure 3b, the metasurface has a bandgap that covers a wide frequency range (~35% bandwidth), which separates two-fold mode degeneracy at K/K′ valleys at both the lower and upper bands. The relatively wider bandgap here, which is a manifestation of synthetic gauge field in the system,[26] is expected due to the stronger interaction between the surface modes over the constitutive layers, which is inversely proportional to the inter-layer spacing. In addition, this case demonstrates the immunity of the proposed dual-metasurface structure to perturbations to the meta-cells design. In general, the edge modes maintain their topological gapless character as long as the frequency detuning between the Dirac cones of the independent TE and TM modes remains smaller than the topological bandgap opened by the magneto-electric coupling.

## 2.3. Realization and Observation of Robust Edge Transport

The stability of edge states based on QSHE, as in the proposed system, is generally warranted by the time-reversal symmetry in the absence of spin-flipping processes. This in addition to the spin-chirality feature of the edge states makes PTIs attractive for making waveguides that are robust against defects and disorder such as sharp bends. To verify the existence of topological edge states in the dual-metasurface and confirm their robustness, a prototype based on standard printed-circuit-board (PCB) technique was fabricated and tested. The prototype is configured such that the orientation of the complementary metallic patterns is inverted in both lateral and vertical directions across the waveguide interface, as shown in Figure 4a (only top layer is visible). The sample has 35µm-thick copper layers on a 1.57mm-thick Rogers/Duroid 5880 substrate with permittivity of 2.2 and tangent loss factor of 0.002. The unit cell size (periodic constant) is 7mm and the width of the slit (strip) of the patch (aperture) is 0.6mm. The design has a center operating frequency of 16.25 GHz, making the total thickness about 1/12 of the wavelength. Note that the substrate thickness is chosen to provide adequate mechanical support but, in principle, thinner design can be realized instead, which would have advantageously wider bandgap as discussed above.

First, we tested the existence of the topological bandgap by exciting bulk modes with an antenna source placed far from the domain wall. Figure 4b clearly reveals the gap spanning the frequency range 13.9 GHz – 18.6 GHz, i.e. 28.9% bandwidth (marked by dashed lines), which corresponds to the topological region of the unit cell's band structure from numerical simulations (see Experimental Section). Next, we tested the presence of the topological edge mode by placing the excitation source at the domain wall and measuring the transmission spectrum. We notice enhanced transmission within the bandgap, which is attributed to the excitation of the edge mode. Here, the transmission is roughly steady throughout the middle of the bandgap and decreases towards the lower and upper bulk bands frequencies. The measured results were found to be in good agreement with the full-wave frequency-domain simulations of the entire sample (see Experimental Section).

Finally, to demonstrate the robustness of the proposed PTI metasurface, we tested waveguides featuring sharp corners of 120° (photographed in Figure 4a) and 90°. The transmission data of the bent waveguides are compared to the straight waveguide case in Figure 4b, which depicts similar values. To directly visualize the edge modes, we scanned the near-field distribution of electric-field intensity over the upper surface ($xy$ plane) of the entire sample area. As shown in Figure 4c, the measured amplitude ratios of $E_z$ show good contrast at 16 GHz (i.e. within the bulk bandgap), with minimum energy away from the domain wall and simultaneously roughly constant amplitude along the zigzag path. This map proves that the wave is indeed guided by the interface line between the two inverted dual-metasurfaces with good localization and negligible scattering losses at the bends. Note that the edge modes are also robust against 90° bends, as demonstrated in Figure 4d, since they could also occur along the armchair edge of the hexagonal metasurface, unlike the so-called valley topological insulators.[44–47]

## 2.4. Discussion

PTIs are preferably robust also against lattice perturbations, i.e. displacements in the local arrangements of unit cells, which are a common uncertainty in photonic crystals fabrication. Since the topological phase in the proposed system stems from EM duality, which is tied to the local response and basic design of the unit cell,[29] it is immune to such perturbations. This is unlike crystalline topological insulator structures,[19–23] in which lattice disorder and improper edge cuts may break topological protection and open a gap in the edge mode spectrum. In general, breaking the underlying symmetry behind the system's topology leads to the removal of essential degeneracies in the band structure. In our system the same would happen if the duality is violated. However, as discussed earlier, the proposed dual-metasurface is insensitive to perturbations to the meta-cells design as long as the spectral detuning between the Dirac cones of the independent TE and TM modes remains smaller than the topological bandgap open by the magneto-electric coupling.

Compared with other PTIs, the proposed metasurface does not require external magnetic field[11,12,15,48] nor bulky structures[14,16–18] for realizing robust edge modes, thus benefiting both large-scale production and applications. Note that the reported design is easily reproducible up to terahertz frequencies,[49,50] where metallic dissipation losses are relatively low. In addition, the planar geometry and ultrathin thickness of the metasurface will facilitate its integration with existing electronic systems. This includes electronic topological insulators as well as PTIs based on lumped-circuit components.[20,51,52] Attractively, the addition of electrical elements to the proposed PCB-compatible design would enable tunable/switchable topological states and reconfigurable pathways.[53] Other elements including mechanical resonators and superconducting Josephson junctions, which are useful for advanced information processing, could also be included.[54] Furthermore, the considered open boundary nature here is attractive due to the possible interaction between topological modes and free-space waves, which paves a way to metasurface designs with unique topologically-endowed scattering characteristics.[55]

Additionally, the demonstrated feasibility for scanning the near fields over the metasurface generally allows more direct experimental studies on topological phenomena of classical waves.

Notably, unlike the majority of existing PTIs, our design approach relies on engineering surface waves rather than bulk waves. Uniquely, the reported edge states are confined along a 1D line rather than a surface interface. Note that common PTIs structures are 2D systems, where their interface modes are either free to travel in two dimensions or have an enclosing structure such as parallel plate waveguide for vertical confinement. This makes the proposed design in comparison appealing for energy confinement and transport applications, as a 1D object being potentially the smallest waveguide possible. Advantageously, this potentially enables topological modes with strong field enhancement, which is beneficial for light-matter interactions. Moreover, the enhanced effective bianisotropy of the dual-metasurface, which stems from the strong coupling between surface waves, creates a bandgap with bandwidth greater than 25% compared to previously reported topological bandgap of less than 10%.[25–31]

## 3. Conclusion

In summary, we have demonstrated an ultrathin topologically nontrivial metasurface based on EM duality, which emulates QSHE and exhibits directional gapless edge modes. Specifically, we have shown how the EM duality underpinning the necessary spin degeneracies is restored by combining complementary metallic patterns of hexagonal symmetry. We have also shown experimentally by near field mapping robust edge states transport through sharp corners with negligible scattering, a critical concern in conventional waveguides. Unlike existing PTIs, our design approach is straightforward and enables broad operating bandwidth and edge states that are notably confined along 1D line interface. This paves the way for planar, compact and efficient routing and concentration of EM energy endowed by topological properties. The reported metasurface is attractive as a tabletop platform for the study of photonic topological phases, as well as for applications benefiting the compactness and versatility of metasurfaces and the potential of topological insulators.

## 4. Experimental Section

*Simulations*: Throughout this paper, all full-wave simulations and numerical calculations are performed using commercial software ANSYS HFSS, in which the eigen-modal and driven-modal setups are used. When calculating the bulk (edge) band structure, periodic boundary conditions are imposed on periodic surfaces of the unit cell (supercell). In case of domain wall simulation, the structure in surrounded by vacuum with sufficient distance comparable to the wavelength at the low frequency end.

*Calculations of Berry Curvature*: When numerically calculating Berry curvature of a point in the reciprocal space, we first integrate Berry connection along an infinitesimal square contour around the point. Then, according to Stokes' theorem, the line integral is equal to the surface integral of Berry curvature over the infinitesimal square that includes the point. Hence, the Berry curvature of the point is the line integral divided by the area of the infinitesimal square. In our numerical calculations, the path integrals were discretized into summations and the side length of the square contour was chosen to be $\delta k = 1/40a$, where $a$ is the period constant.

*Measurement Set-up*: A near-field 2D scanning system was used for measurements. A probe antenna centered at the interface line and oriented in the direction of propagation of the edge mode was used as the excitation source. Another probe, which is oriented vertically to the surface, was used to scan the relative intensity of normal component of electric-field ($E_z$) in close proximity to the top surface of the PCB board. In measurements, the source and detector are connected to port 1 and port 2 of a vector network analyzer and the scattering parameter S21 is recorded, which is proportional to the $E_z$. The scan resolution in the *xy* plane is $0.5 \times 0.5 \text{mm}^2$. The total area of the fabricated dual-metasurfaces on both sides of the domain wall is $18 \times 24 \text{ cm}^2$.

**Conflict of Interest**

The Authors declare no competing financial interests.

**Acknowledgements**

This work has been supported in part by AFOSR grant FA9550-16-1-0093, and by DARPA contract W911NF-17-1-0580.

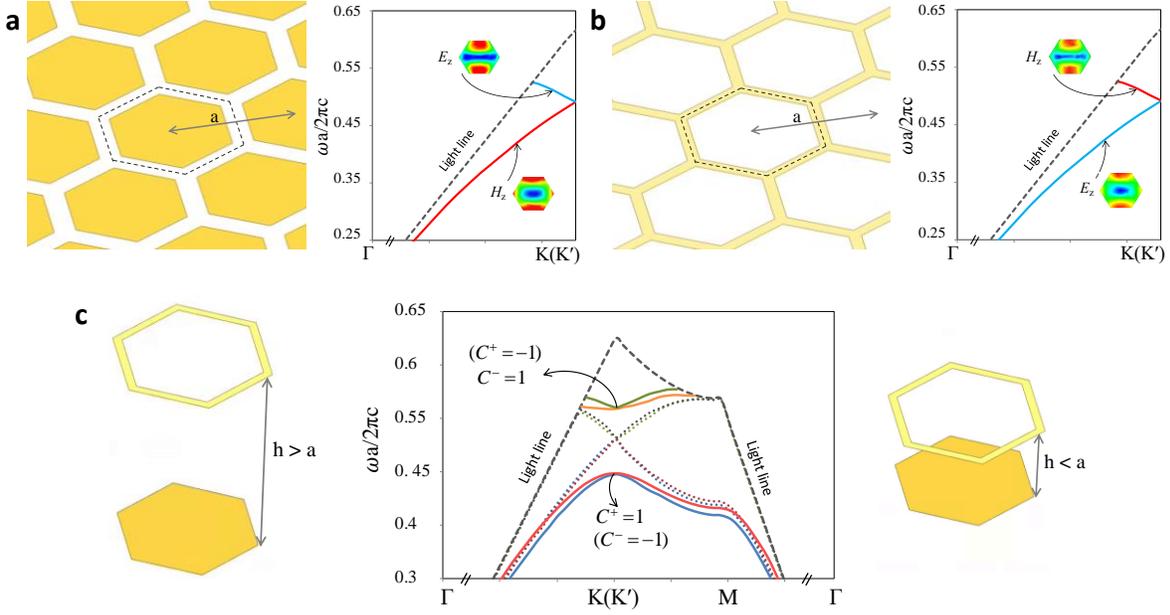

**Figure 1.** Unit cell design of the dual-metasurface and corresponding band structure. (a) schematic of the patch-type metasurface and the associated band structure. The unit cell is outlined by dashed black line. (b) schematic of the aperture-type metasurface and the associated band structure. (c) schematic of the EM dual-metasurface and the corresponding band structure. Bands in the case of large inter-layer separation are marked with dotted lines whereas the bands in the case of small inter-layer separation are marked with solid lines. $C^+$ and $C^-$ refer to spin-Chern number for $\psi^+$ and $\psi^-$ states, respectively. The shown values without (within) parentheses indicate the non-trivial topological invarient is due to Berry phase accumelation around K (K′) point.

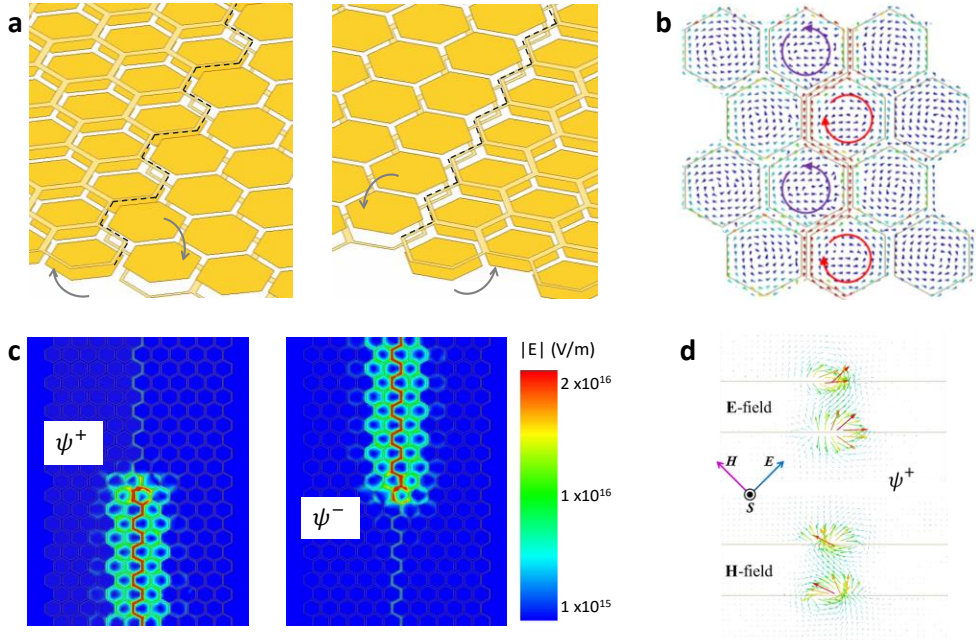

**Figure 2.** Edge states between topological dual-metasurfaces with opposite signs of effective bianisotropy. (a) schematic of the domain wall between dual-metasurfaces with inverted orientation in the normal (z) direction. Choice of either armchair (left) and zigzag (right) edge boundary is appropriate. (b) In-plane power flux (pointing vector) on both sides of the domain wall illustrating the orbital state of the spin-up mode propagating in the forward direction. (c) selective excitation of the spin-polarized edge mode with appropriate electric and magnetic dipole sources showing spin-momentum locking. (d) electric and magnetic vector fields distributions over a cross-section of the domain wall showing high localization of the edge mode along the line interface between the inverted dual-metasurfaces.

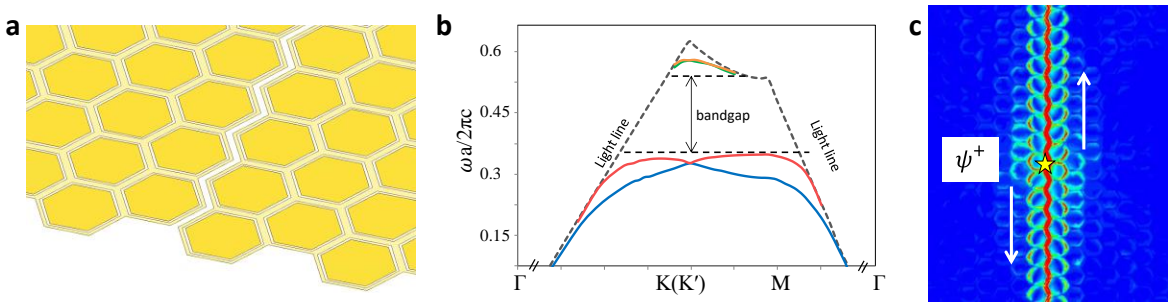

**Figure 3.** Edge states between topologically nontrivial (dual-metasurface) and trivial (air) domains. (a) schematic of metasurface-air interface made using two identical flat dual-metasurfaces placed side by side with a finite gap in between. (b) the band structure of the flat dual-metasurface's unit cell on either side. Large topologically nontrivial bandgap is present despite the patch-layer and aperture-layer not being perfectly complementary to each other. (c) planar surface field distribution across the slot showing excellent localization of the edge modes. The spin state is excited here unidirectionally due to the opposite orientation of the metasurface-air interface across the slot.

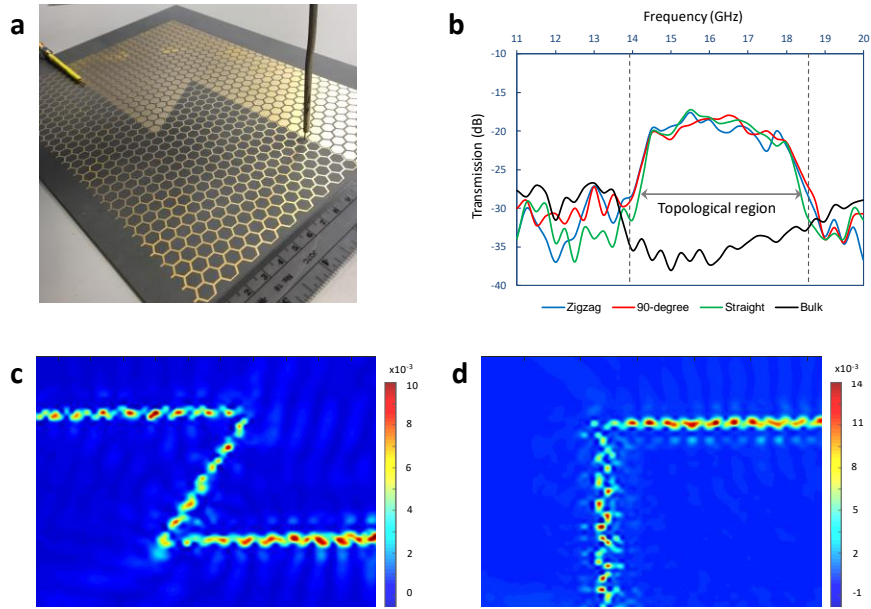

**Figure 4.** Measured transmission of robust edge states and direct observation via near field mapping of surface field distribution. (a) Photograph of the fabricated prototype of the dual-metasurface PTI waveguide, which is made by interfacing two structures with opposite sign of bianisotropy, along with the excitation and detection antennas. (b) transmission data of bulk and edge modes showing the existence of bulk bandgap and the enhanced transmission of edge modes within the bandgap. Here, similar transmission values are observed for straight, zigzag, and 90° waveguides, verifying the robustness of the PTI against sharp bends. (c) measured $E_z$ field distribution at 16 GHz over the surface of a PTI with zigzag interface showing localization of the edge mode at the domain wall and the successful transmission of the edge mode with roughly constant field intensity. (d) measured $E_z$ field distribution at 16 GHz over the surface of a PTI with sharp 90°corner showing similar localization and transmission features to the zigzag case.

**ToC:**

This work proposes a new approach to realize photonic topological insulators without the restrictive requirements and complex designs of existing systems. This is done by coupling surface waves between two overlapped ultrathin metasurfaces of complementary properties. This enables a vanishingly-narrow line interface, which supports direction-specific modes. These modes are experimentally shown to propagate over a wide bandwidth through sharp corners with negligible scattering, a critical concern in conventional waveguides.

**Keyword:** Topological metasurfaces

D. J. Bisharat*, D. F. Sievenpiper

**Title:** Electromagnetic-dual metasurfaces for topological states along a one-dimensional interface

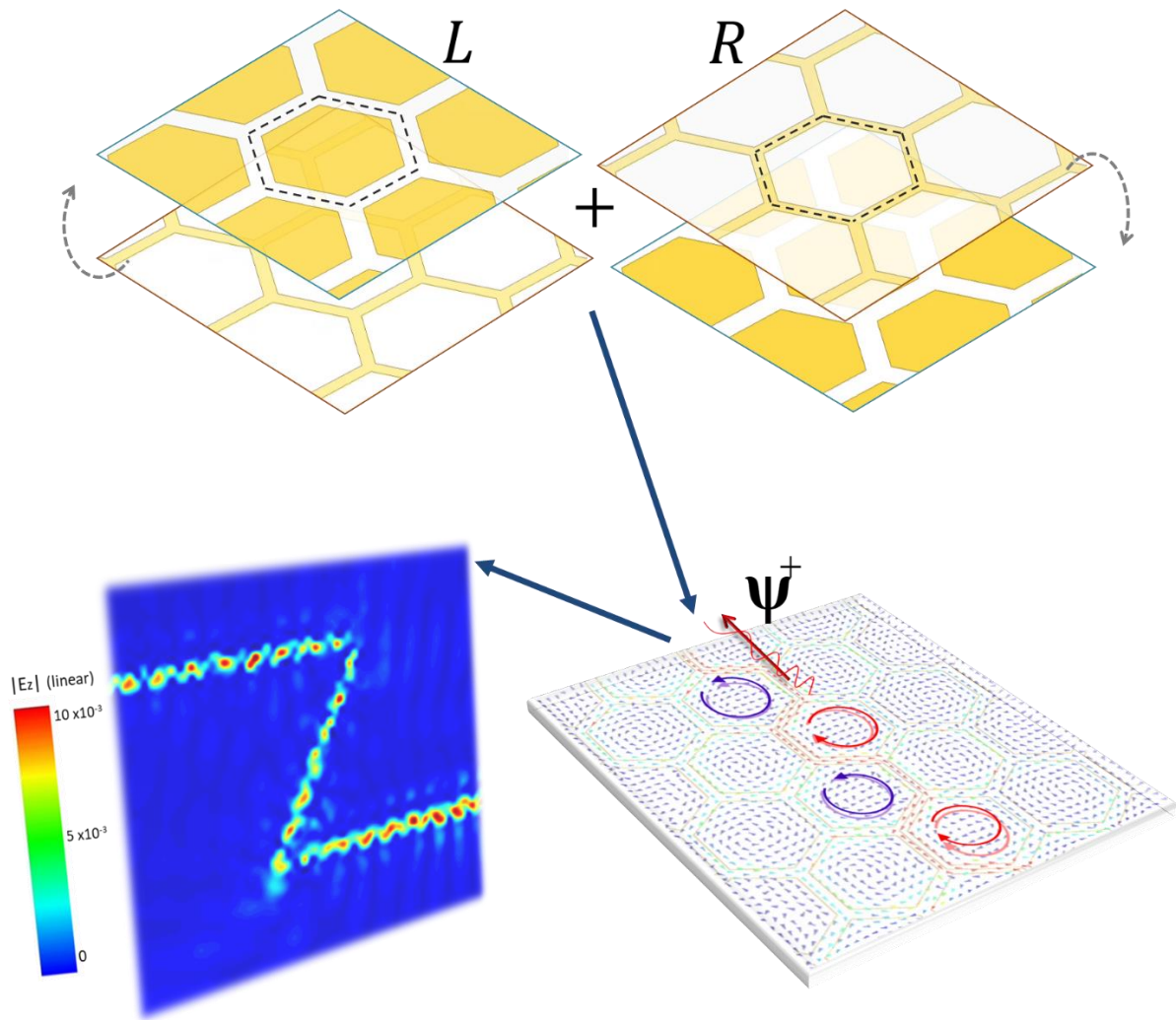